\documentclass[review]{elsarticle}

\usepackage{lineno,hyperref}

\usepackage[ngerman]{babel}
\usepackage{graphicx}
\usepackage{fancybox}
\usepackage{shadow}
\usepackage{amsmath,amssymb,bm}
\usepackage{epsfig}
\usepackage{mdwlist}
\usepackage{amsfonts}
\usepackage{setspace}
\usepackage{rotating}
\usepackage{graphics}
\usepackage{fancybox}
\usepackage{shadow}
\usepackage{amsmath,amssymb}
\usepackage{amsfonts}
\usepackage{accents}
\usepackage{color}
\usepackage{setspace}
\usepackage{tabularx}
\usepackage{dcolumn}
\usepackage{lscape}
\usepackage{bm} 
\usepackage{longtable}
\usepackage{multirow}
\addto\captionsngerman{
    
    } 

\journal{Journal of Molecular Spectroscopy}

\begin{document}

\begin{frontmatter}

\title{Accidental Interactions in the Torsional Spectrum of Singly Deuterated Hydrogen Peroxide HOOD\\
\vspace{0.4 cm}
\normalsize{In memoriam Dr. Jon T. Hougen}
}

\author{D. Herberth$^1$, T.F. Giesen$^1$}
\address{$^1$University of Kassel, Heinrich-Plett-Str. 40, D-34132 Kassel, Germany\\}
\fntext[myfootnote]{Corresponding author: D. Herberth\\
\textit{Email address}: herberth@physik.uni-kassel.de\\
\textit{URL}: www.uni-kassel.de/go/labastro}

\author{K.M.T. Yamada$^2$}
\address{$^2$ National Institute of Advanced Industrial
Science and Technology, Tsukuba-West,
Onogawa 16-1, Tsukuba, Ibaraki 305-8569, Japan\\}





\begin{abstract}
We extended previously recorded infrared spectra of singly deuterated hydrogen peroxide (HOOD) to the submm-wavelength region and derived accurate molecular parameters and a semi-empirical equilibrium structure. 
In total, more than 1500 ro-torsional HOOD transitions have been assigned between 6 and 120 cm$^{-1}$. 
We succeeded to analyze the accidental interaction between the torsional sub-states by measuring several perturbed transitions. In addition to the set of Watsonian parameters for each tunneling component, only two interaction constants were required to describe the spectrum. The $K_a$- and $J$-dependance of the torsional splitting could be determined also for the perturbed states.\\
\end{abstract}


\end{frontmatter}



\section{Introduction} 

Hydrogen peroxide H$_2$O$_2$ with its torsional tunneling is a well-known role-model for large amplitude motion in molecules \cite{Hougen84} and the simplest case study of a chiral molecule showing internal rotation. 
In its  ground state, H$_2$O$_2$ shows a skew-chain structure with $C_2$ symmetry \cite{Penny}, while its deuterated form HOOD has no geometrical symmetry ($C_1$). As in the case of HSOH \cite{Yamada1}, the two hydrogen peroxide enantiomers are separated by a \textit{trans}- and \textit{cis}-barrier, respectively. \cite{Hunt}.\\
H$_2$O$_2$ plays an important role in the chemistry of the earth's stratospheric ozone layer and was also detected in a star forming region towards $\rho$ Ophiuchi A \cite{Bergman}. 
The torsion-rotation spectrum of HOOH has extensively been studied in the past in the microwave and infrared region, as early as 1934 Penny and Sutherland succeeded to derive the HOOH ground state structure from their recorded spectra (\cite{Penny}). In 1950 Giguere \textit{et al.} \cite{Giguere} recorded extended infrared spectra to obtain molecular parameters which were in agreement with the predictions made by Penny and Sutherland.\\
Ever since then, experimental evidence has been accumulated from many sources. More references to the studies of the torsion-rotation spectrum of hydrogen peroxide can be found in our previous publication \cite{Herberth2012} and for example in Camy-Peyret \textit{et al.} \cite{Camy-Peyret}.\\
In 2001, Bak \textit{et al.} found a large discrepancy in structural parameters for HOOH when comparing calculated and empirical molecular structures, originating in a lack of precise experimental data on the hydrogen peroxide isotopomers \cite{Bak}.
In the same year, Koput \textit{et al.} \cite{Koput} noted the lack of precise experimental data on the hydrogen peroxide isotopomers and calculated high-quality spectroscopic constants for H$_{2}$O$_{2}$, HOOD, D$_{2}$O$_{2}$ and H$_{2}\!^{18}$O$_{2}$ to support a future analysis of the rotation-torsion spectra of these molecules.\\ After Flaud \textit{et al.} \cite{Flaud2} determined experimental rotational constants for DOOD, we tried to close the final gap by publishing experimental rotational constants for HOOD in 2012 \cite{Herberth2012}. But due to perturbations between crossing torsional states, which could not be analysed with the dataset available to us at that time, the rotational constants were preliminary.\\
 Despite its asymmetric structure HOOD is a nearly prolate symmetric top with Ray’s asymmetry parameter of $\kappa =0.985$. 
The OH and OD group undergo a torsional motion about the OO bond, which is hindererd by a barrier in the $trans$- and $cis$-position, respectively \cite{Flaud}.
Tunneling through these two barriers causes a splitting of energy levels in four torsional sub-levels, denoted by $\tau_1$, $\tau_2$, $\tau_3$ and $\tau_4$.\\
Due to symmetry reasons, for the main species HOOH only states with even numbered $K_a$ exist in the torsional substates $\tau_1$ and $\tau_4$, while in $\tau_2$ and $\tau_3$ only odd $K_a$ values can be found. In contrast to this, in the case of HOOD the pseudo-quantum number $\tau$ loses its clear meaning to distinguish the torsional sub-levels in relation with the $cis$ and $trans$ tunnelling barrier. Additionally, the large \textit{cis}-barrier in HOOD (2460 cm$^{-1}$ \cite{Hunt}) leads to a negligibly small \textit{cis}-splitting of energy levels. It is useful for our analysis to merge the labels $\tau_1$ and $\tau_2$ to one torsional sub-level, which is denoted by $v_\mathrm{LAM}=0$ (LAM stands for large amplitude motion), and in analogy $\tau_3$ and $\tau_4$ are merged to $v_\mathrm{LAM}=1$.\\
We found these two sets of torsional levels to be seperated by a splitting of 5.786(13) cm$^{-1}$ \cite{Herberth2012}, determined by the \textit{trans}-barrier height (381 cm$^{-1}$ \cite{Hunt}), the reduced-mass of the torsional motion and on the rotational quantum numbers $J$ and $K_a$.\\
Due to its torsional motion, HOOD is an ideal testbed for recent quantum chemical models describing internal rotation. HSOH, which like HOOD has $C_1$ symmetry, shows an alternating size of the torsional splitting with $K_a$ modulo 3. This effect could be explained by a model by Hougen \textit{et al.} \cite{Hougen84} and Yamada \textit{et al.} \cite{Yamada1}, in which the energy splitting is expressed in terms of \textit{cis}- and \textit{trans}-tunneling interaction. 
In our previous publication we could not observe any $K_a$-alternation in the size of the torsional splitting and due to a limited dataset at low frequencies and perturbations in that region, also the assignment of low $K_a$ levels was not possible.\\
In this work, we measured high resoluted microwave gas phase spectra of HOOD between 180 - 800 GHz in our Kassel THz laboratory, including several perturbed transitions in low $K_a$ levels. 
In combination with our previous far infrared measurements of the molecule, we were able to derive highly accurate experimental rotational constants for the ground state.
In addition to one set of Watsonian parameters for each tunneling doublet component, only two interaction constants are required to describe the perturbation and to reproduce the more than 1500 assigned transitions within the experimental uncertainty. The $K_a$- and $J$-dependance of the torsional splitting could be determined also for the perturbed states. On the basis of the rotational constants we calculated a semi-empirical equilibrium structure of HOOH by taking into account \textit{ab initio} zero-point vibrational corrections.\\ 


\section{Experiment}
As described by Flaud \textit{et al.} \cite{Flaud2} a sample of HOOD was prepared by mixing a 30 percent dilute solution of hydrogen peroxide with the same amount of D$_2$O and evaporating the liquid under reduced pressure to increase the concentration of HOOD. More detailed information concerning the sample preparation is given in our previous publication \cite{Herberth2012}.
The resulting solution contained H$_2$O$_2$, HOOD
and D$_2$O$_2$ with a ratio of 1:2:1. The sample gas was continuously injected into the gas cell, maintaining a slow and constant flow with total pressure between 0.05 and 2 mbar. 

The infrared spectra of HOOD were recorded at the SOLEIL synchrotron facility during 15 shifts of beamtime at the AILES Beamline, making use of a Bruker IFS 125 Fourier Transform Infrared Spectrometer and a 40 m White-type multireflexion cell. The beamline properties and the application to high resolution gas-phase spectroscopy are described in details in Refs. \cite{Brubach,Cuisset,McKellar}, more details concerning the measurement conditions are given in \cite{Herberth2012}.

To identify formerly unassigned perturbed transitions, which lie in lower frequency regions unaccessible to SOLEIL, we employed the high resolution THz spectrometer at Kassel University to record submillimeter spectra between 180 and 800 GHz (6.1 - 25.9 cm$^{-1}$). The radiation was generated by a microwave synthesizer equipped with an amplifier-multiplier chain and was detected by a zero-bias Schottky diode. To improve the signal-to-noise-ratio, the FM-modulated signal was demodulated by a Lock-In-amplifier. All measurements were performed at room temperature in a 3.5 m long absorption cell under gas flow conditions and at a total pressure of 0.05 - 0.1 mbar.\\


\section{Data Analysis}
In total, more than 1500 transitions of the molecule were assigned to the vibrational ground state in the frequency range of 6 to 120 cm$^{-1}$ up to $J=30$ and $K_{a}=11$, as listed in Table \ref{Table maxJ}. The assignment procedure is described in detail in Ref. \cite{Herberth2012}.\\

\begin{table}[h]
\begin{center}
\caption{Maximum values of $J$ observed in each branch of HOOD\label{Table maxJ}}
\begin{tabular}{c|ccc|ccc|ccc|ccc}
\hline
\\
&  \multicolumn{3}{c}{$\Delta v_{\mathrm{LAM}}=1 \leftarrow 0$}	
&  \multicolumn{3}{c}{$\Delta v_{\mathrm{LAM}}=0 \leftarrow 1$} 
&  \multicolumn{3}{c}{$\Delta v_{\mathrm{LAM}}=0 \leftarrow 0$} 
&  \multicolumn{3}{c}{$\Delta v_{\mathrm{LAM}}=1 \leftarrow 1$} \\
\\
\hline
\\
$K_{\mathrm{a}}''$ 
& $^rP$ 
& $^rQ$ 
& $^rR$ 
& $^rP$ 
& $^rQ$ 
& $^rR$ 
& $^rP$ 
& $^rQ$ 
& $^rR$ 
& $^rP$ 
& $^rQ$ 
& $^rR$ 
\\
0	&	2  &	24&	21&	-  &	- & 5 &	-  &	18  & -  & -  & 19& - \\
1	&	-  &	12&	23&	-  &	11&	19&	-  &	-  &	-  &	-  &	13&	28\\
2	&	-  &	15&	27&	-  &	22&	13&	-  &	-  &	27&	-  &	22&	29\\
3	&	-  &	20&	27&	-  &	21&	22&	-  &	-  &	22&	-  &	-  &	20\\
4	&	19&	28&	22&	14&	28&	20&	-  &	12&	19&	-  &	16&	17\\
5	&	17&	30&	18&	-  &	27&	19&	-  &	17&	19&	-  &	22&	17\\
6	&	13&	24&	11&	16&	30&	17&	-  &	20&	16&	-  &	22&	14\\
7	&	22&	28&	9  &	16&	25&	10&	-  &	20&	13&	-  &	30&	13\\
8	&	21&	22&	-  &	27&	27&	8  &	14&	-  &	-  &	11&	-  &	-\\
9	&	25&	-  &	-  &	23&	-  &	-  &	12&	-  &	-  &	18&	-  &	-\\
10	&	-  &	-  &	-  &	20&	-  &	-  &	-  &	-  &	-  &	27&	-  &	-\\
11	&	28&	-  &	-  &	25&	-  &	-  &	-  &	-  &	-  &	-  &	-  &	-\\
\\
\hline
\hline
\end{tabular}
\end{center}
\end{table}

We used Colin Western`s software PGOPHER \cite{pgo} for the least squares fit analysis, employing a standard Watson type Hamiltonian in S reduction. The two components $v_{\mathrm{LAM}}=0$ and $v_{\mathrm{LAM}}=1$ were treated as two separated states.
In this manner the $J$- and $K$-dependences of the tunneling splitting are effectively taken care of as the differences of the rotational paremeters for the $v_{\mathrm{LAM}}=0$ and 1 states.\\
The preliminarily obtained molecular constants suggested that there should two avoided level crossings be observable in our accessible frequency range: 
One between the $v_{\mathrm{LAM}}=0$ and $v_{\mathrm{LAM}}=1$ components of $K_a=1$ level at $J$ = 16 
and another one between the $v_{\mathrm{LAM}}=0$ and $v_{\mathrm{LAM}}=1$ components of $K_a=2$ level at $J$ = 29.
Note, that a similar avoided crossing was observed for HSOH $K_a=2$ levels \cite{Ross}.\\
With the help of our new THz measurements and with careful and tedious comparison of the simulated and observed spectra, we succeeded to assign a number of perturbed transitions.\\
We found that the perturbations can be described by centrifugal-correction terms: $\hat{h} = W_{ac}(\hat{J}_a\hat{J}_c + \hat{J}_c\hat{J}_a)/2 + W_{bc} (\hat{J}_b\hat{J}_c + \hat{J}_c\hat{J}_b)/2$, where the interaction constants $W_{ac}$ and $W_{bc}$ contain the
displacement coordinate of the internal rotation. In addition to the set of Watsonian parameters for each tunneling component, the interaction constants between the tunneling pairs are determined very accurately by the least-squares procedure provided in PGOPHER. All 41 parameters of the fit are given in Table \ref{ABCTabelle}.\\

The final fit has a root-mean-square error (rms) of 0.0004 cm$^{-1}$, lying well within the experimental uncertainty of 0.001 cm$^{-1}$ but slightly higher than the 0.00036 cm$^{-1}$ in our previous analysis \cite{Herberth2012}. The main reason for this increase lies in the assignment of many new lines in the low frequency region of the SOLEIL data. The lower quality of the SOLEIL measurements at low frequencies leads to lower frequency accuracies of the assigned transitions, resulting in a larger rms.
In the present fitting, a standard deviation of 540 kHz was chosen for the THz data, while a standard deviation of 14 MHz was selected for the SOLEIL data, resulting in a dimensionless standard deviation of 0.996, which indicates that the weights used are reasonable.\\
Compared to our previously published parameters \cite{Herberth2012}, we observed a change of one to three orders of magnitude in the off diagonal Quartic and Sextic Centrifugal Distortion constants $d_1$, $d_2$, $h_1$ and $h_2$ for the $v_\mathrm{LAM}$= 0 level, while the change in these parameters for the $v_\mathrm{LAM}$= 1 level is much smaller. In addition, these constants are better determined by one to two orders of magnitude. The main reason lies in the assignment of several transitions in the $K_a=1$ $v_\mathrm{LAM}=0$ state, which now determine the size of the $K$-type-doubling for this level, mainly described by the above-mentioned parameters. On the contrary, there was already a sufficient amount of transitions assigned for $K_a$= 1 $v_\mathrm{LAM}=1$, thus the $K$-type-doubling for this state did not undergo a significant change in our new analysis.\\ 
As a consequence, the rotational constants $A$, $B$ and $C$ for the $v_\mathrm{LAM}=0$ level are determined with higher accuracy, which now enables us to derive an empirical equilibrium molecular structure of hydrogen peroxide, as is described in section \ref{Struktur}.\\
The interaction parameters $W_{ac}$ and $W_{bc}$ will be discussed in more detail in the next paragraph.\\ 

\begin{table*}[h!]
\begin{center}
\caption{Experimentally determined molecular constants of HOOD for the ground
state. Uncertainties in parentheses are 1$\sigma$ from the
least square fit in units of the last quoted digit.\label{ABCTabelle}} 
\begin{tabular}{cr@{}lr@{}lr} 
\\
\hline
Parameter 
& \multicolumn{2}{c}{$v_{\mathrm{LAM}}=0$}   
& \multicolumn{2}{c}{$v_{\mathrm{LAM}}=1$ }  
& Unit\\
\hline
$A$		&	212836&.710(70)		&	212665&.337(63)	&MHz\\
$B$ 		&	24848&.380(35)		&	24812&.967(32)	&MHz\\
$C$ 		&	23397&.985(36)		&	23447&.915(34)	&MHz\\
$D_J$		&	82&.16(33)		&	79&.31(33)	&kHz\\	
$D_{JK}$	&    	0&.9507(11)		&	0&.9715(10)	&MHz\\
$D_K$		&       6&.79919(33)		&	6&.7085(34)	&MHz\\
$d_1$		&    	-89&.05(32)		&	-1&.072(17) 	&kHz\\
$d_2$		&     	-5&.430(42)		&	-2&.694(21) 	&kHz\\
$H_J$		&      	5&.34(84)		&      	4&.23(82)  	&Hz\\
$H_{JK}$	&  	-35&.3(36) 		&    	-20&.1(35)   	&Hz\\
$H_{KJ}$	&    	0&.507(16)		&     	0&.572(15) 	&kHz\\
$H_{K}$		&     	1&.136(29)		&     	1&.102(38) 	&kHz\\
$h_1$		&      -1&.503(49)		&   	-1&.620(29) 	&Hz\\
$h_2$		&       0&.709(65)		&    	1&.144(58) 	&Hz\\
$h_3$		&      -1&.284(28)		&  	-31&.4(25)   	&Hz\\
$L_J$		&      -9&.07(60)		&    	-8&.56(58)	&mHz\\
$L_{JJK}$	&    	10&.5(33)		&     	6&.2(33) 	&mHz\\
$L_{JK}$	&     	-1&.778(23)		&  	-0&.948(24)	&Hz\\
$L_{KKJ}$	&    	-4&.774(73)		&  	-3&.06(10)	&Hz\\
$\Delta E$	&	&
\multicolumn{2}{r}{173461.54(20)~~~~}  &  	&MHz\\
$|W_{ac}|$ 	& 	& \multicolumn{2}{r}{210.38(82)~~~~} 	& 	&MHz\\
$W_{bc}$ 	& 	& \multicolumn{2}{r}{171.5014(18)}	& 	&MHz\\
\hline
\hline
\end{tabular}
\end{center}
\end{table*}


\subsection{Accidental resonance between torsional tunneling pairs}

\begin{figure}[h]
\fbox{\includegraphics[width=12cm] {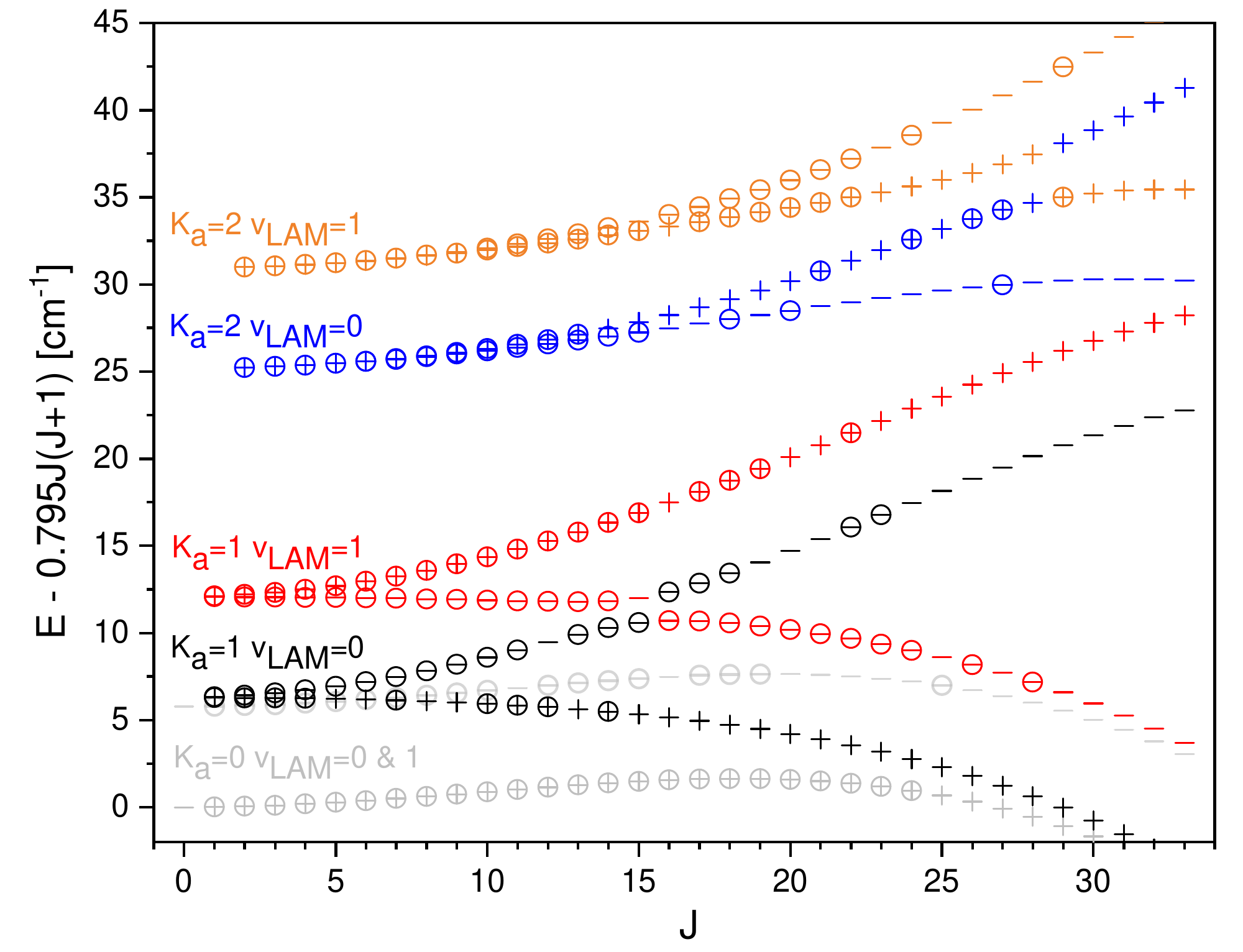}}
\caption{HOOD energy levels in the torsional ground state up to $K_a = 2$ and $J=34$. $K_a = 0$ levels are indicated only lightly in order to avoid complication. The plus and minus signs indicate the parity of states. Data points that are based on measured transitions are marked by circles. The accidental resonances occur between the $K_a=1$, $v_\mathrm{LAM}=0$ and $K_a=1$, $v_\mathrm{LAM}=1$ levels, as well as between the $K_a=2$, $v_\mathrm{LAM}=0$ and $K_a=2$, $v_\mathrm{LAM}=1$ levels.}
\label{fig:elevels}
\end{figure}

In the present study, we observed two accidental resonances between torsional tunneling pairs of HOOD: One between the $v_{\mathrm{LAM}}=0$ and $v_{\mathrm{LAM}}=1$ components of the $K_a=1$ level at approximately $J$ = 16 and another one between the $v_{\mathrm{LAM}}=0$ and $v_{\mathrm{LAM}}=1$ components of the $K_a=2$ level at approximately $J$ = 29, see Fig. \ref{fig:elevels}.

\begin{figure} [h!]
\begin{center}
\fbox{\includegraphics[width=11cm] {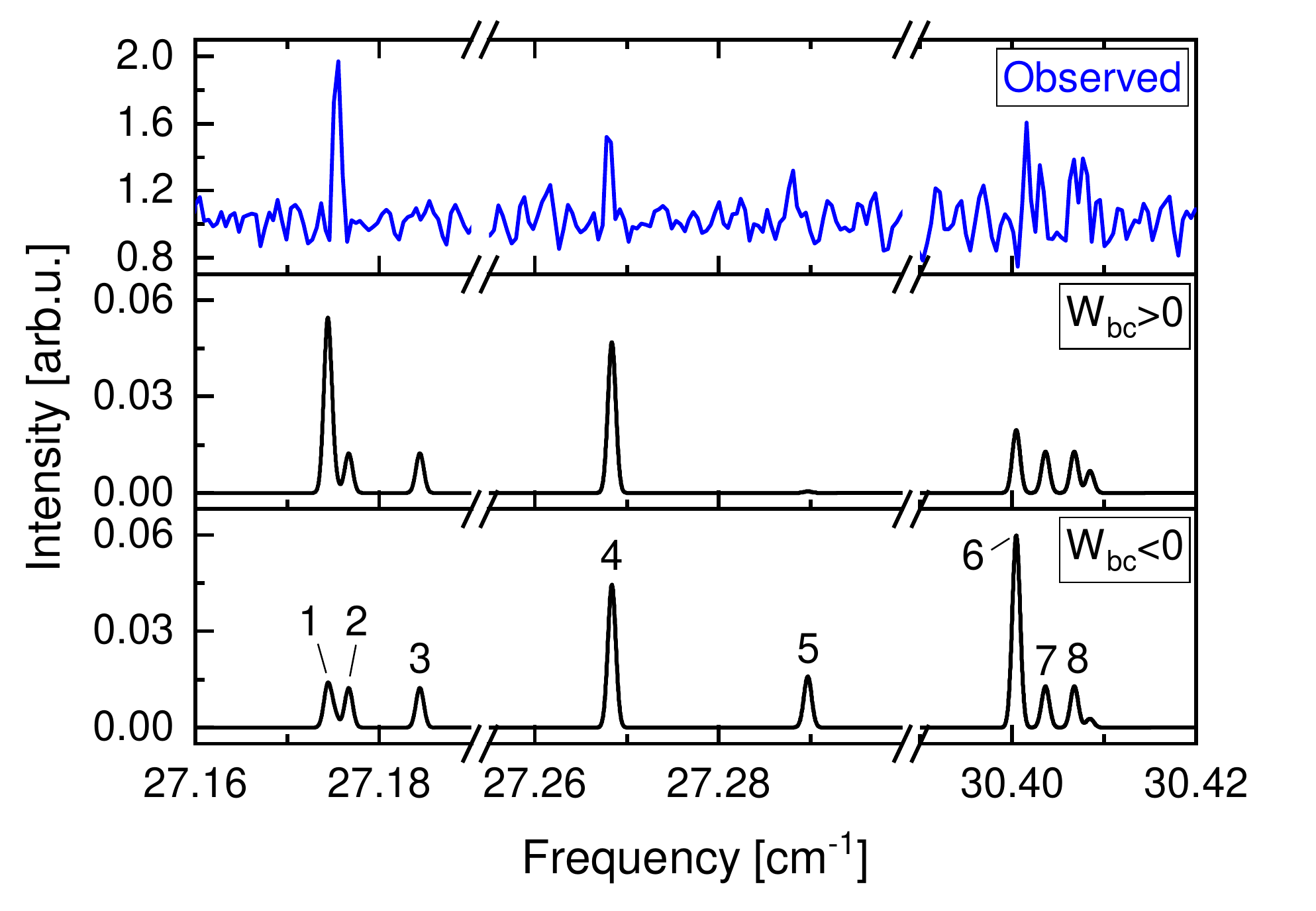}}
\end{center}
\caption{Comparision of the observed spectrum with simulated spectra with different signs of $W_{bc}$. The sign of $W_{ac}$ has shown no significant influence on the spectra. For transition quantum numbers please refer to Table \ref{tab:check}.}
\label{fig:intwbc2}
\end{figure}

These resonances can be described by centrifugal-correction terms of the form:
\begin{equation}
\hat{h} = W_{ac}(\hat{J}_a\hat{J}_c + \hat{J}_c\hat{J}_a)/2 + W_{bc} (\hat{J}_b\hat{J}_c + \hat{J}_c\hat{J}_b)/2
\end{equation}
where the interaction constants $W_{ac}$ and $W_{bc}$ contain the displacement coordinate of the 
internal rotation. The perturbation selection rules for the first term ($ac$-coupling) and the second term ($bc$-coupling) are different from each other: \\
$ac$-coupling: $\Delta K_a = \pm 1$\\
$bc$-coupling: $\Delta K_a = 0, \pm 2$\\
And for both couplings applies: $\Delta v_{LAM} = \pm 1$ and $\text{parity:} \; \pm \longleftrightarrow \pm$\\
It can easily be shown that from the fitted line positions only the absolute values of $W_{ac}$ and $W_{bc}$ can be derived. To additionally determine the signs of the two perturbation constants, we have to consider also the line intensities. Thus, we have compared the relative observed and simulated intensities for several lines for four different cases: \\
(a)		$W_{ac}>0, W_{bc}>0$  ;         (b)	$W_{ac}>0, W_{bc}<0$	\\
(c)		$W_{ac}<0, W_{bc}>0$  ;         (d)	$W_{ac}<0, W_{bc}<0$     \\
For simulation by PGOPHER we used the dipole moments $\mu_a=0.0458$ D, $\mu_b=0.5593$ D, and $\mu_c=1.7252$ D estimated by \textit{ab initio} calculations \cite{Herberth2012}.
The results are summarized in Table \ref{tab:check} and depicted in Fig. \ref{fig:intwbc2}.

\begin{table*}

\caption{Intensities are checked for selected transitions of HOOD.}\label{tab:check} 
\begin{center}
\begin{tabular}{c|ll|r|c|c|c|c}
\hline
& Upper & Lower & Position & \multicolumn{4}{c}{Relative intensity agreement}\\
& Level & Level &\multirow{2}{*}{[cm$^{-1}$]}& $W_{ac}>0$&$W_{ac}>0$&$W_{ac}<0$&$W_{ac}<0$\\
&\multicolumn{2}{c|}{$v_\mathrm{LAM}$, $J_{K_a,K_c}$}&&$W_{bc}>0$&$W_{bc}<0$&$W_{bc}>0$&$W_{bc}<0$\\
\hline
1)& $0,15_{1,14}$ &$1,14_{0,14}$  & 27.1744  &      &     &      &    \\
2)& $1,13_{4,9}$ &$0,14_{3,11}$   & 27.1766  & good & bad & good & bad\\
3)& $1,13_{4,10}$ &$0,14_{3,12}$  & 27.1843  &      &     &      &    \\
\hline
4)& $1,13_{2,11}$ &$0,13_{1,13}$  &27.2683  &\multirow{2}{*}{good$^{(a)}$}&\multirow{2}{*}{bad}&\multirow{2}{*}{good}&\multirow{2}{*}{bad}\\
5)& $1,19_{0,9}$ &$1,18_{1,18}$   &27.2897  &      &     &      &     \\
\hline
6)& $0,16_{1,15}$ &$1,15_{0,15}$   &30.4004  &     &     &      &     \\
7)& $1,13_{4,10}$ &$0,14_{3,12}$   &30.4036  & good& bad & good & bad \\
8)& $1,11_{4,8}$ &$0,12_{3,10}$    &30.4067  &     &     &      &     \\
\hline
\multicolumn{8}{l}{{\footnotesize $^{(a)}$ The absorption peak near 27.288 cm$^{-1}$ in the observed spectrum is not line 5 in this table}}\\
\hline
\hline
\end{tabular}
\end{center}
\end{table*}

The sign of $W_{ac}$ has shown to have no significant influence on the simulated spectra so it could so far not be determined. The intensity change due to the perturbation by the $W_{ac}$ term may be too small to detect in the observed lines.\\
The two crossing levels $K_a$ = 1, v$_{LAM}$ = 0 and $K_a$ = 1, v$_{LAM}$ = 1 are interacting by $bc$-coupling. Since we assigned a significant amount of transitions in these two levels near the crossing point, the $W_{bc}$ constant could be determined with high accuracy, while the $W_{ac}$ constant, for which no crossing levels could be assigned in proximity to a crossing point, shows lower accuracy.\\ 

\newpage


\subsection{Torsional splitting}

\begin{figure}[h!]
\begin{center}
\fbox{\includegraphics[width=11cm] {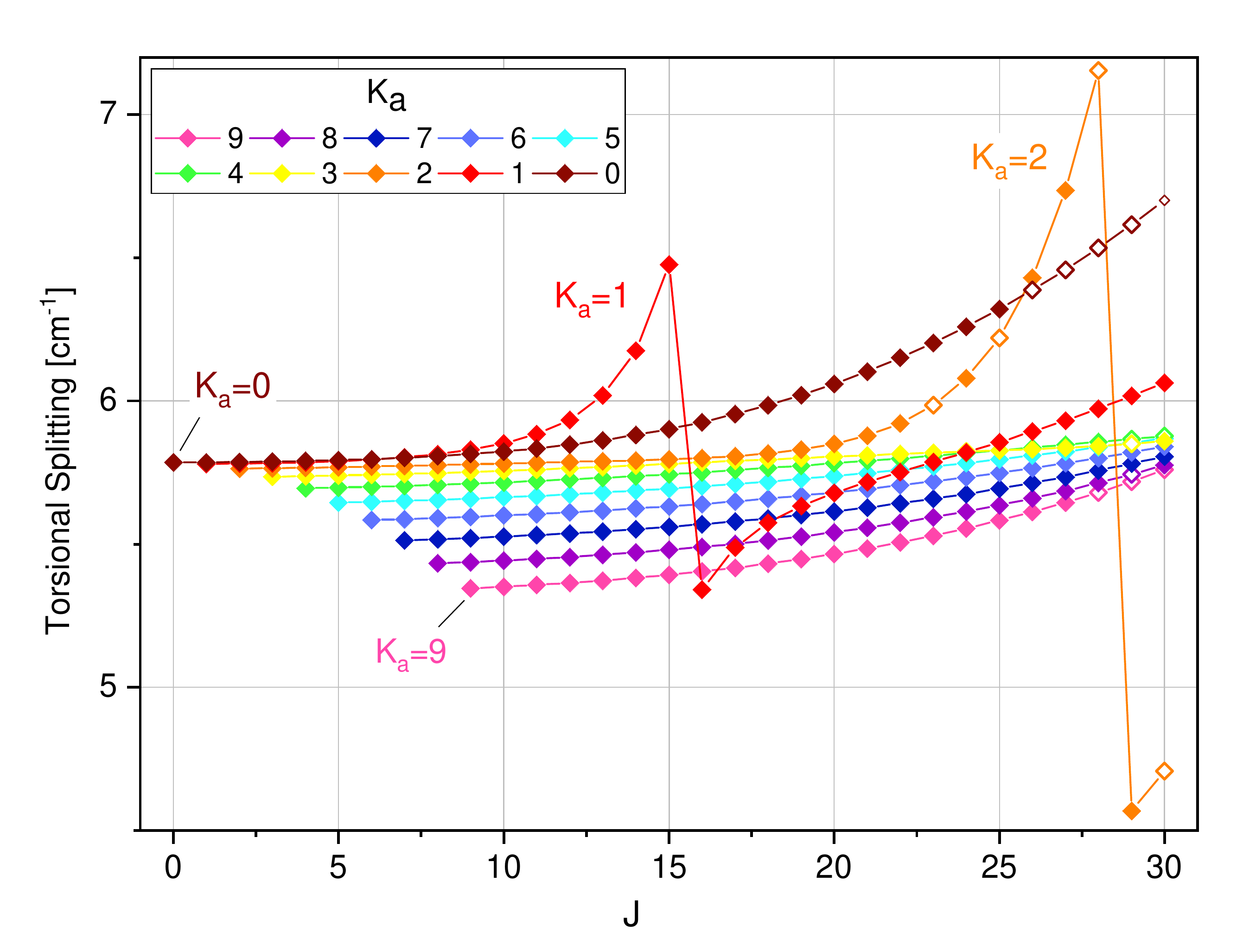}}
\end{center}
\caption{Size of the torsional splitting for different $K_a$ in dependance of $J$. Filled symbols: Based on experimental data, blank symbols: Calculated data}
\label{fig:tauJ}
\end{figure}

\begin{figure}[h!]
\begin{center}
\fbox{\includegraphics[width=9cm] {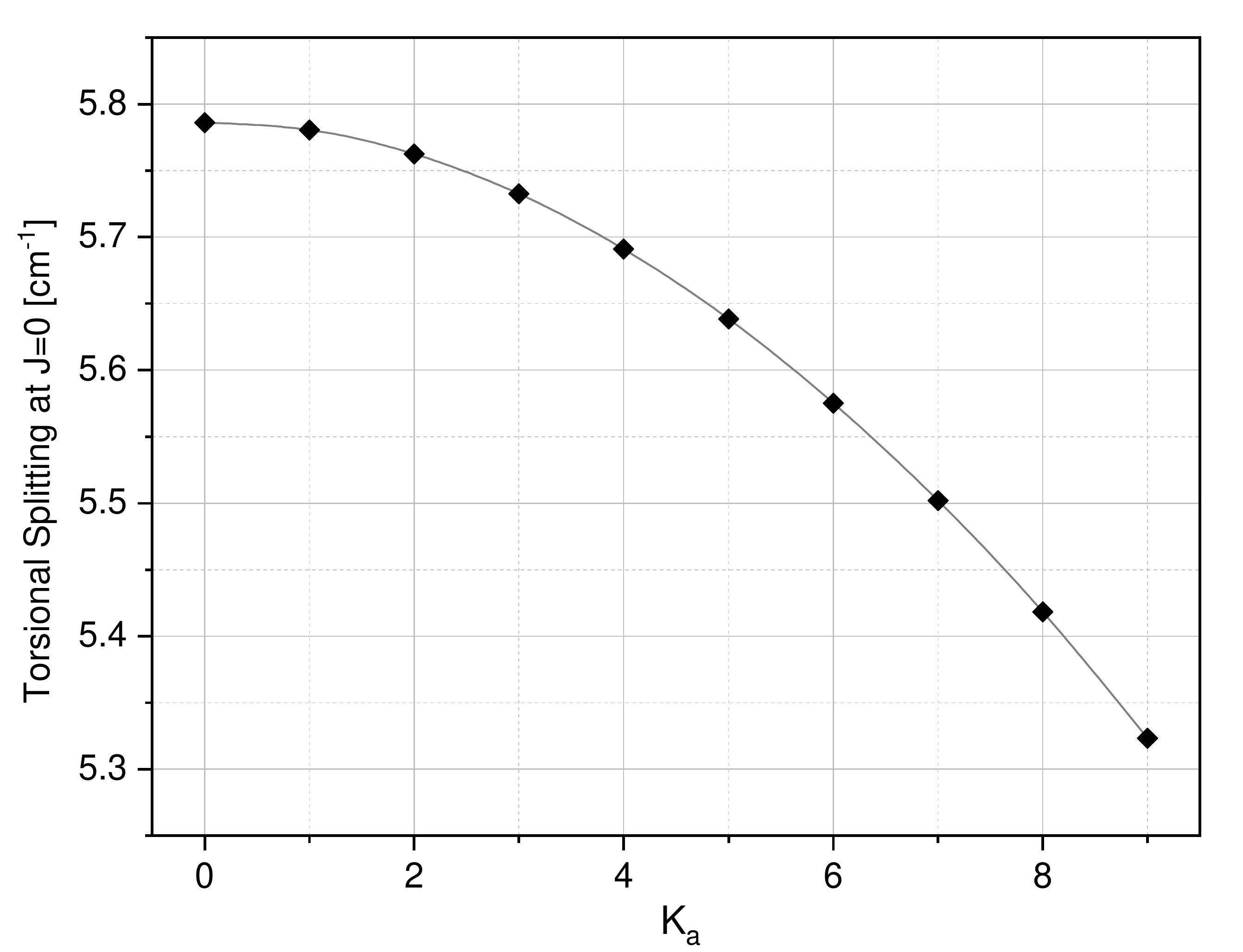}}
\end{center}
\caption{Size of the torsional splitting, extrapolated to $J$=0, in dependance of $K_a$. }
\label{fig:tauKa}
\end{figure}

HOOD undergoes a torsional motion, hindered by a $trans$- and a $cis$-barrier. Tunneling through these barriers causes a splitting of energy levels depending on the molecular symmetry, the barrier heights, the reduced mass of the torsional motion and the rotational quantum numbers $J$ and $K_a$ \cite{Baum}, \cite{Jabs}.\\
Like HOOD, HSOH has no geometrical symmetry and shows nearly the same ratio of moments of inertia of the rotating moieties (SH, OH and OH, OD respectively). Baum \textit{et al.} \cite{Baum} observed an alternation of the size of the torsional splitting for HSOH with a period of approximately modulo $K_a=3$, which can be explained by the interaction between $cis$- and $trans$-tunneling \cite{Yamada1}.\\
But unlike in the case of HSOH, no $K_a$-alternation of the torsional splitting could be observed for HOOD in our previous work \cite{Herberth2012}, in which we determined the torsional splitting with the help of combination differences of \textit{b}- and \textit{c}-type transitions.\\
Consequently, in this work, we calculate the torsional splitting directly from the energy levels which were exported from PGOPHER \cite{pgo}, resulting from the least squares fit of the 39 rotational constants and the two perturbation constants (Table \ref{ABCTabelle}) to the 1500 assigned transitions. After averaging over the two $K_a$-type doubling components for each $K_a$, $v_\mathrm{LAM}$ level in the first step, we afterwards calculated the difference between $v_\mathrm{LAM}=1$ and $v_\mathrm{LAM}=0$ for each $K_a$ and $J$.\\
By calculating the torsional splitting from the energy levels, we can take advantage of our larger dataset of assigned transitions - especially at low $K_a$ values and in form of perturbed transitions - and we can now determine the size of the torsional splitting for a larger range of $J$ and $K_a$, as depicted in Fig. \ref{fig:tauJ}.\\
As can clearly be seen for $K_a=1$ and 2, the two avoided crossings of energy levels manifest in characteristic deviations of the torsional splitting. These deviations are larger for $K_a=2$, since the crossing in this case takes place at higher values of $J$ and the centrifugal correction term contains $J^2$.\\
The $K_a$ dependance of the torsional splitting (Fig. \ref{fig:tauKa}) was determined by fitting polynomial functions to the curves of Fig. \ref{fig:tauJ} and extrapolating them to $J=0$.\\
The decrease of the torsional splitting with $K_a$ indicates that the torsional motion is accompanied by skeletal flexing of HOOD. Since higher rotational excitations stabilize the molecule, the splitting collapses.\\
In contrast to HSOH, where the \textit{trans}- and \textit{cis}-barrier have comparable sizes ($V_{trans}= 1520.9$ cm$^{-1}$ and $V_{cis}= 2163.3$ cm$^{-1}$ \cite{Yurchenko}), 
the HOOD \textit{cis}-barrier is more than six times larger than the \textit{trans}-barrier (2460 cm$^{-1}$ and 381 cm$^{-1}$ \cite{Hunt}).\\
Apparently, the contribution of the \textit{cis}-splitting to the total torsional splitting is too small to cause an observable alternation effect in the case of HOOD.\\
\newpage


\section{Semi-Empirical Equilibrium Structure of H$_2$O$_2$} \label{Struktur}

We derived a semi-empirical equilibrium structure of hydrogen peroxide on the basis of our empirically determined rotational constants \textit{A$_{0}$}, \textit{B$_{0}$}, and \textit{C$_{0}$} of HOOD together with previously published empirical rotational constants of D$_{2}$O$_{2}$ \cite{Flaud2} and H$_{2}$O$_{2}$ \cite{Petkie98}, see Table \ref{ABCalphaTabelle}.\\
Semi-experimental equilibrium rotational constants were obtained from the relation $B_e=B_0-\frac{1}{2}\sum_{r}\alpha^{b}_{r}$ with similar equations holding for $A_0$ and $C_0$.
The zero-point vibrational corrections $\frac{1}{2}\sum_{r}\alpha^{a,b,c}_{r}$ to the rotational constants for all normal modes were computed at the fc-CCSD\-(T)\-/cc-pVQZ level using a perturbative scheme as outlined by Stanton \textit{et al.} \cite{Stanton}.\\
The empirical equilibrium structure of H$_2$O$_2$ was derived from a least-squares fit of the OH and OO bonds, the HOO bond angle $\delta$ and the HOOH dihedral angle $\tau$ to the equilibrium moments of inertia of the normal and two isotopic species, on the assumption that the molecule exhibits $C_2$ symmetry.\\
The resulting structural parameters are given in Table \ref{StrukturParameter}, along with a semi-empircal structure by Baraban \textit{et al.}. 
For comparison reasons, we also included a high level \textit{ab initio} calculated structure by Malyszek \textit{et al.} \cite{Malyszek} as well an experimental structure derived by Pelz \textit{et al.} \cite{Pelz}.\\
Baraban determined semi-experimental equilibrium rotational constants by subtracting \textit{ab initio} rotation-vibration zero-point corrections $\frac{1}{2}\sum_{r}\alpha_{r}$ from experimentally observed rotational constants. These corrections were obtained by perturbation treatment of the nuclear motion on potential energy surfaces. 
The structure by Baraban listed in Table \ref{StrukturParameter} was determined by the same procedure as described in his former publication  \cite{Baraban}, but with our newest rotational constats for HOOD (Table \ref{ABCTabelle}).\\ 
Our experimental structure parameters are in excellent agreement with those of Baraban within errors, with the only exception of the OOH angle, but with a marginal deviation of 0.08 degrees.\\

\begin{table*}[h]
\caption{Ground state rotational constants and vibrational corrections for HOOH and its isotopomers in MHz. $\frac{1}{2}\sum_{r}\alpha_{r}$ are the vibration rotation interaction constants for all normal modes r. \label{ABCalphaTabelle}} 
\begin{center}
\begin{tabular}{ l | r@{.}l r@{.}l }
\hline
 Parameter  & \multicolumn{2}{l}{Emp. value}	& \multicolumn{2}{c}{$\frac{1}{2}\sum_{r}\alpha_{r}^{(a)}$}\\
\hline
  HOOH$^{(b)}$      		\\
  $A$       & 301730&3501 &   2679&1145  \\
  $B$       &  26168&4074 &    314&9160  \\
  $C$       &  25158&1076 &    443&4456  \\
  DOOD$^{(c)}$                        \\
  $A$       & 165084&5121 &    895&4465  \\
  $B$       &  23328&7325 &    200&6947  \\
  $C$       &  22018&5001 &    337&0233  \\
  HOOD$^{(d)}$                      \\
  $A$       & 212751&0163   &   1477&210   \\
  $B$       &  24830&6800   &    255&652   \\
  $C$       &  23422&9588  &    381&625   \\
\hline
\multicolumn{5}{l}{{\footnotesize $^{(a)}$ This work, fc-CCSD(T)/cc-pVQZ.}}\\
\multicolumn{5}{l}{{\footnotesize $^{(b)}$ Petkie \textit{et al.} \cite{Petkie98}, $^{(c)}$ Flaud \textit{et al.} \cite{Flaud}}}\\
\multicolumn{5}{l}{{\footnotesize $^{(d)}$ This work, average rotational constants}}\\
\multicolumn{5}{l}{{\footnotesize \hspace{0.4cm} determined for $v_\mathrm{LAM}=0$ and 1}}\\
\hline
\hline
\end{tabular}
\end{center}
\end{table*}

\begin{table} [h!]
\caption{Geometrical equilibrium parameters for the vibrational ground state of hydrogen peroxide. Uncertainties in parentheses are $1\sigma$ from the least-squares fits in units of the last quoted digit, distances are given in $\mathring{A}$, angles in degrees.}\label{StrukturParameter}
\begin{center}
\begin{tabular}{l|cccc}
\hline
Parameter	&	{Baraban \textit{et al.}$^{(a)}$}	&	{This work}	& 	{Malyszek \textit{et al.} } & Pelz \textit{et al.}\\
			&	{semi emp.}		&	{semi emp.}	&	{\cite{Malyszek} \textit{ab initio}}& \cite{Pelz} (exp.) \\
\hline
  $r_{e}$(OO)      	 &  1.4523(2) 	&  1.4518(6) &  1.4509  & 1.4556 \\
  $r_{e}$(OH)        &  0.9618(2) 	&  0.962(1)  &  0.9628  & 0.967  \\
  $\delta_{e}$(HOO)  &  99.78(4) 	&  100.0(1)  &  100.09  & 102.32 \\
  $\tau_{e}$(HOOH)   &  113.5(2) 	&  113.7(2)  &  112.81  & 113.70 \\
\hline
\multicolumn{4}{l}{{\footnotesize $^{(a)}$unpublished results}}\\
\hline
\hline
\end{tabular}
\end{center}
\end{table}


\newpage

\section{Conclusions}

The gas phase spectrum of HOOD was measured in the microwave region and combined with infrared data from an earlier study, in total more than 1500 ro-torsional transitions could be assigned in the vibrational ground state. 
By adding newly assigned perturbed transitions to the existing data base, we derived more precise rotational constants, which enabled us to determine a semi-empirical equilibrium structure of hydrogen peroxide, which is in agreement with recently published semi-empirical values by Baraban \textit{et al.} \cite{Baraban}.\\
The size of the torsional splitting in dependance of $J$ and $K_a$ was determined, also for the perturbed states at low $K_a$ values, and no $K_a$ alternation of the latter could be observed. Refering to the models describing internal rotation splittings \cite{Hougen84}, this is due to the different height of \textit{trans}- and \textit{cis}- barrier in HOOD.\\
The precise rotational constants for HOOD now allow for assignment of torsionally excited transitions, which are included in our measured far infrared dataset. The analysis of the first torsionally excited state is in progress, a publication will follow.\\
Winnewisser \textit{et al.} \cite{WinnewisserHSOH} observed a similar kind of interaction between torsional sub-states in the spectra of HSOH earlier. The present analysis of the HOOD spectra, however, strongly suggests that the interaction analysis reported in Ref. \cite{WinnewisserHSOH} for HSOH might be inadequate. Thus, we are planning to reanalyze the spectra of HSOH.\\


\section{Acknowledgements}

This work has been funded by the Deutsche Forschungsgemeinschaft (DFG, German Research Foundation) – Projektnummer 328961117 – SFB 1319 ELCH. We thank Dr. S. Thorwirth for the calculation of \textit{ab initio} vibrational corrections. SOLEIL beamtime was allocated under project 20090856. D. Herberth thanks Dr. U. Hildebrandt, Dr. C. Clemens, Dr. F. Schmidt and their great team for their kind support. 



\newpage

\section*{Appendix}
\renewcommand{\arraystretch}{0.7} 
\setlength{\tabcolsep}{1.3mm}
\begin{longtable}{cccccccccccc}
\caption{List of newly measured HOOD transitions. Standard Deviation: 0.54 MHz} \\
No&Obs&Cal&Obs-Cal&$v_\mathrm{LAM}'$&$J'$&$K_a'$&$K_c'$&$v_\mathrm{LAM}''$&$J''$&$K_a''$&$K_c''$\\
\hline
\endfirsthead 
\caption{Continuation of newly measured HOOD transitions}\\
No&Obs & Calc & Obs-Calc &	$v_\mathrm{LAM}'$	&	$J'$	&	$K_a'$	&	$K_c'$	&	$v_\mathrm{LAM}''$	&	$J''$	&	$K_a''$	&	$K_c''$	\\
\hline
\endhead 
\multicolumn{4}{r}{Continuation on the next page}\\
\endfoot
\hline
\multicolumn{4}{r}{End of table} \\
\endlastfoot
1)	&	189211.372	&	189211.806	&	-0.434	&	1	&	1	&	1	&	0	&	1	&	1	&	0	&	1	\\
2)	&	189433.029	&	189432.737	&	0.292	&	0	&	1	&	1	&	0	&	0	&	1	&	0	&	1	\\
3)	&	190585.706	&	190586.456	&	-0.75	&	1	&	2	&	1	&	1	&	1	&	2	&	0	&	2	\\
4)	&	190891.394	&	190891.298	&	0.096	&	0	&	2	&	1	&	1	&	0	&	2	&	0	&	2	\\
5)	&	192663.553	&	192664.598	&	-1.045	&	1	&	3	&	1	&	2	&	1	&	3	&	0	&	3	\\
6)	&	193091.399	&	193091.492	&	-0.093	&	0	&	3	&	1	&	2	&	0	&	3	&	0	&	3	\\
7)	&	195464.635	&	195465.896	&	-1.261	&	1	&	4	&	1	&	3	&	1	&	4	&	0	&	4	\\
8)	&	196048.102	&	196048.295	&	-0.193	&	0	&	4	&	1	&	3	&	0	&	4	&	0	&	4	\\
9)	&	199015.235	&	199016.557	&	-1.321	&	1	&	5	&	1	&	4	&	1	&	5	&	0	&	5	\\
10)	&	199781.109	&	199781.306	&	-0.197	&	0	&	5	&	1	&	4	&	0	&	5	&	0	&	5	\\
11)	&	203347.855	&	203348.939	&	-1.084	&	1	&	6	&	1	&	5	&	1	&	6	&	0	&	6	\\
12)	&	204313.576	&	204313.665	&	-0.089	&	0	&	6	&	1	&	5	&	0	&	6	&	0	&	6	\\
13)	&	208500.322	&	208501.052	&	-0.731	&	1	&	7	&	1	&	6	&	1	&	7	&	0	&	7	\\
14)	&	209670.225	&	209670.115	&	0.11	&	0	&	7	&	1	&	6	&	0	&	7	&	0	&	7	\\
15)	&	214515.7	&	214516.007	&	-0.308	&	1	&	8	&	1	&	7	&	1	&	8	&	0	&	8	\\
16)	&	215873.9	&	215873.626	&	0.274	&	0	&	8	&	1	&	7	&	0	&	8	&	0	&	8	\\
17)	&	221441.7	&	221441.44	&	0.26	&	1	&	9	&	1	&	8	&	1	&	9	&	0	&	9	\\
18)	&	222939.6	&	222939.217	&	0.383	&	0	&	9	&	1	&	8	&	0	&	9	&	0	&	9	\\
19)	&	229329.7	&	229328.862	&	0.838	&	1	&	10	&	1	&	9	&	1	&	10	&	0	&	10	\\
20)	&	230862.25	&	230861.794	&	0.456	&	0	&	10	&	1	&	9	&	0	&	10	&	0	&	10	\\
21)	&	241354.18	&	241354.331	&	-0.152	&	1	&	24	&	1	&	24	&	0	&	24	&	0	&	24	\\
22)	&	245035.12	&	245035.133	&	-0.013	&	1	&	23	&	1	&	23	&	0	&	23	&	0	&	23	\\
23)	&	248211.68	&	248210.353	&	1.327	&	1	&	12	&	1	&	11	&	1	&	12	&	0	&	12	\\
24)	&	249125.3	&	249125.168	&	0.132	&	1	&	22	&	1	&	22	&	0	&	22	&	0	&	22	\\
25)	&	253566.17	&	253566.016	&	0.153	&	1	&	21	&	1	&	21	&	0	&	21	&	0	&	21	\\
26)	&	258268.12	&	258267.793	&	0.327	&	1	&	20	&	1	&	20	&	0	&	20	&	0	&	20	\\
27)	&	258563.37	&	258563.507	&	-0.137	&	0	&	13	&	1	&	12	&	0	&	13	&	0	&	13	\\
28)	&	259320.36	&	259319.097	&	1.263	&	1	&	13	&	1	&	12	&	1	&	13	&	0	&	13	\\
29)	&	263086.68	&	263086.394	&	0.286	&	1	&	19	&	1	&	19	&	0	&	19	&	0	&	19	\\
30)	&	267316.04	&	267316.468	&	-0.428	&	0	&	14	&	1	&	13	&	0	&	14	&	0	&	14	\\
31)	&	267770.84	&	267770.565	&	0.275	&	1	&	18	&	1	&	18	&	0	&	18	&	0	&	18	\\
32)	&	271617.49	&	271616.718	&	0.772	&	1	&	14	&	1	&	13	&	1	&	14	&	0	&	14	\\
33)	&	271829.53	&	271829.388	&	0.142	&	1	&	17	&	1	&	17	&	0	&	17	&	0	&	17	\\
34)	&	273092.94	&	273093.338	&	-0.398	&	0	&	15	&	1	&	14	&	0	&	15	&	0	&	15	\\
35)	&	274222.81	&	274222.91	&	-0.1	&	1	&	16	&	1	&	16	&	0	&	16	&	0	&	16	\\
36)	&	285159.04	&	285158.989	&	0.052	&	1	&	15	&	1	&	14	&	1	&	15	&	0	&	15	\\
37)	&	312987.85	&	312987.804	&	0.046	&	1	&	14	&	1	&	14	&	0	&	14	&	0	&	14	\\
38)	&	315130.06	&	315129.856	&	0.204	&	1	&	13	&	1	&	13	&	0	&	13	&	0	&	13	\\
39)	&	316179.93	&	316181.12	&	-1.19	&	1	&	17	&	1	&	16	&	1	&	17	&	0	&	17	\\
40)	&	319290.91	&	319290.69	&	0.22	&	1	&	12	&	1	&	12	&	0	&	12	&	0	&	12	\\
41)	&	324263.08	&	324262.876	&	0.204	&	1	&	11	&	1	&	11	&	0	&	11	&	0	&	11	\\
42)	&	329487.05	&	329486.829	&	0.221	&	1	&	10	&	1	&	10	&	0	&	10	&	0	&	10	\\
43)	&	333746.81	&	333747.579	&	-0.769	&	1	&	18	&	1	&	17	&	1	&	18	&	0	&	18	\\
44)	&	334672.76	&	334672.539	&	0.221	&	1	&	9	&	1	&	9	&	0	&	9	&	0	&	9	\\
45)	&	336700.23	&	336701.065	&	-0.835	&	0	&	17	&	1	&	16	&	0	&	17	&	0	&	17	\\
46)	&	339645.39	&	339645.117	&	0.273	&	1	&	8	&	1	&	8	&	0	&	8	&	0	&	8	\\
47)	&	344285.5	&	344285.159	&	0.341	&	1	&	7	&	1	&	7	&	0	&	7	&	0	&	7	\\
48)	&	348504.51	&	348504.078	&	0.431	&	1	&	6	&	1	&	6	&	0	&	6	&	0	&	6	\\
49)	&	352233.55	&	352233.002	&	0.548	&	1	&	5	&	1	&	5	&	0	&	5	&	0	&	5	\\
50)	&	352729.73	&	352728.239	&	1.491	&	1	&	19	&	1	&	18	&	1	&	19	&	0	&	19	\\
51)	&	353381.69	&	353381.562	&	0.128	&	0	&	18	&	1	&	17	&	0	&	18	&	0	&	18	\\
52)	&	355417.998	&	355417.35	&	0.648	&	1	&	4	&	1	&	4	&	0	&	4	&	0	&	4	\\
53)	&	358014.72	&	358013.999	&	0.72	&	1	&	3	&	1	&	3	&	0	&	3	&	0	&	3	\\
54)	&	359990.574	&	359989.689	&	0.885	&	1	&	2	&	1	&	2	&	0	&	2	&	0	&	2	\\
55)	&	361320.94	&	361320.088	&	0.852	&	1	&	1	&	1	&	1	&	0	&	1	&	0	&	1	\\
56)	&	531327.18	&	531328.496	&	-1.316	&	1	&	10	&	2	&	8	&	1	&	10	&	1	&	9	\\
57)	&	536972.6	&	536973.213	&	-0.613	&	1	&	9	&	2	&	7	&	1	&	9	&	1	&	8	\\
58)	&	542272.46	&	542273	&	    -0.541	&	1	&	8	&	2	&	6	&	1	&	8	&	1	&	7	\\
59)	&	547156.72	&	547156.996	&	-0.276	&	1	&	7	&	2	&	5	&	1	&	7	&	1	&	6	\\
60)	&	551562.57	&	551562.683	&	-0.113	&	1	&	6	&	2	&	4	&	1	&	6	&	1	&	5	\\
61)	&	555435.71	&	555435.994	&	-0.284	&	1	&	5	&	2	&	3	&	1	&	5	&	1	&	4	\\
62)	&	558730.43	&	558731.272	&	-0.842	&	1	&	4	&	2	&	2	&	1	&	4	&	1	&	3	\\
63)	&	561410.43	&	561411.117	&	-0.687	&	1	&	3	&	2	&	1	&	1	&	3	&	1	&	2	\\
64)	&	563445.52	&	563446.14	&	-0.62	&	1	&	2	&	2	&	0	&	1	&	2	&	1	&	1	\\
65)	&	567534.31	&	567534.597	&	-0.287	&	1	&	2	&	2	&	1	&	1	&	2	&	1	&	2	\\
66)	&	569566.64	&	569566.665	&	-0.025	&	1	&	3	&	2	&	2	&	1	&	3	&	1	&	3	\\
67)	&	572275.62	&	572275.493	&	0.127	&	1	&	4	&	2	&	3	&	1	&	4	&	1	&	4	\\
68)	&	575660.22	&	575659.495	&	0.725	&	1	&	5	&	2	&	4	&	1	&	5	&	1	&	5	\\
69)	&	579715.88	&	579715.048	&	0.832	&	1	&	6	&	2	&	5	&	1	&	6	&	1	&	6	\\
70)	&	584435.33	&	584434.788	&	0.542	&	1	&	7	&	2	&	6	&	1	&	7	&	1	&	7	\\
71)	&	589804.44	&	589804.582	&	-0.142	&	1	&	8	&	2	&	7	&	1	&	8	&	1	&	8	\\
72)	&	595798.37	&	595797.823	&	0.547	&	1	&	9	&	2	&	8	&	1	&	9	&	1	&	9	\\
73)	&	602364.16	&	602363.915	&	0.245	&	1	&	10	&	2	&	9	&	1	&	10	&	1	&	10	\\
74)	&	609402.8	&	609403.067	&	-0.267	&	1	&	11	&	2	&	10	&	1	&	11	&	1	&	11	\\
75)	&	613716.030	&	613715.996	&	0.0345	&	1	&	5	&	1	&	4	&	0	&	4	&	0	&	4	\\
76)	&	616705.34	&	616705.96	&	-0.62	&	1	&	12	&	2	&	11	&	1	&	12	&	1	&	12	\\
77)	&	623798.02	&	623798.487	&	-0.467	&	1	&	13	&	2	&	12	&	1	&	13	&	1	&	13	\\
\end{longtable}

\end{document}